\documentclass{article}

\usepackage[english]{babel}

\usepackage[letterpaper,top=2cm,bottom=2cm,left=3cm,right=3cm,marginparwidth=1.75cm]{geometry}

\usepackage{array}
\usepackage{amsmath}
\usepackage{graphicx}
\usepackage{tabularx}
\usepackage{csquotes}
\graphicspath{ {Figures/} }  
\usepackage[colorlinks=true, allcolors=blue]{hyperref}
\usepackage[style=ieee,sorting=none]{biblatex}
\title{Blockchain for Trust and Reputation Management in Cyber-physical Systems}

\author{Guntur Dharma Putra$^{1,2}$, Volkan Dedeoglu$^{3}$, Salil S Kanhere$^{1,2}$ and Raja Jurdak$^{4}$}

\date{\small
    $^1$UNSW Sydney $^2$Cyber Security CRC $^3$CSIRO Data61 $^4$QUT Brisbane
}

\begin{document}
\maketitle

\begin{abstract}
The salient features of blockchain, such as decentralisation and transparency, have allowed the development of Decentralised Trust and Reputation Management Systems (DTRMS), which mainly aim to quantitatively assess the trustworthiness of the network participants and help to protect the network from adversaries. In the literature, proposals of DTRMS have been applied to various Cyber-physical Systems (CPS) applications, including supply chains, smart cities and distributed energy trading. In this article, we outline the building blocks of a generic DTRMS and discuss how it can benefit from blockchain. To highlight the significance of DTRMS, we present the state-of-the-art of DTRMS in various field of CPS applications. In addition, we also outline challenges and future directions in developing DTRMS for CPS.
\end{abstract}

\section{Introduction}
\label{sec:intro}
Trust is a subjective and intangible belief built up from consecutive interactions with an entity or individual that perceives its behaviour~\cite{Batwa2021}. As it is linked to a specific behaviour or traits, trust is context-related and thus cannot be generalised. According to Gambetta, trust is defined as a subjective probability that an individual expects from another individual on performing an expected action~\cite{gambetta2000can}. Occasionally, trust and reputation are referred interchangeably in the literature. However, there is a subtle difference between these two terms. Trust refers to a subjective belief towards the behaviour of an entity that builds as more interactions happen, while reputation can be seen as the aggregated opinion or trust degree of an entity from other entities that have prior interaction with the entity.

Trust and Reputation Management Systems (TRMS) aim to assess the accountability or trustworthiness of each participant in distributed systems by means of a quantitative approach. In TRMS, trustworthiness is derived from direct experience or recommendations from other peers and is represented as numerical scores using which the trustworthiness level can be conveniently measured. In general, the trust and reputation score can be used as a safeguard to manage the associated risk in communicating with other peers in a distributed system, which might be very dynamic and hostile.

There have been many applications of TRMS in Cyber-Physical Systems (CPS) and Internet of Things (IoT). For example, TRMS is deployed in the context of social IoT, which is used to assess the trustworthiness of each participating node in the network~\cite{Chen2016}. In e-commerce sites, TRMS is implemented to help customers determine the credibility of the sellers in the marketplace.

However, several challenges exist in building a TRMS. For instance, traditional TRMS architectures rely on a centralised actor to manage the collection of feedback and calculation of trust scores, which raises the risk of data loss and manipulation by the centralised party. When the centralised actor is compromised, an adversary may maliciously alter the trust computation thus undermining the use of these metrics. In addition, authentication and identification of users in TRMS may expose the actual identities of the users, which should be concealed and protected.

Blockchain, the underpinning technology behind Bitcoin, has seen a lot of interest, due to its inherent characteristics, such as traceability, tamper-resilience, trustless environment, programmability, immutability and transparency. These characteristics show promise in addressing the aforementioned issues of TRMS. For instance, blockchain may replace the trusted centralised actor that  assesses the trustworthiness of participants in traditional TRMS. The adoption of blockchain in TRMS is referred to in the literature as Decentralised TRMS (DTRMS)~\cite{Bellini2020}. While blockchain enables trustless interaction between participants, a reputation system is still required to provide some degree of trust quantification for off-chain operations.

In this article, we present and discuss how blockchain technology can be incorporated in TRMS for enhancing its effectiveness for CPS applications. We outline the building blocks of a generic TRMS that delineate how trust is empirically built up by collecting and aggregating evidence of direct and indirect interactions to obtain a quantifiable trust measure. Then, we describe the blockchain properties that can help address challenges in building TRMS for CPS applications. The article also highlights the latest developments of DTRMS for CPS by presenting some recent implementations of DTRMS across various CPS application domains. We also outline the challenges and future directions for DTRMS that still need to be addressed.

The rest of the article is organised as follows. Section~\ref{sec:decentralised-trs} presents the concept of DTRMS. Section~\ref{sec:use-cases} outlines several implementations of DTRMS for CPS. We discuss the open challenges for future research directions in Section~\ref{sec:challenges} and give a conclusion of our article in Section~\ref{sec:conclusion}.

\section{Blockchain-based Trust and Reputation Management Systems}
\label{sec:decentralised-trs}
In this section, we discuss the necessity of a TRMS for CPS along with the general properties of CPS applications, including trust derivation, types of trust, evidence aggregation approach and trust dimensions. We also illustrate the salient properties of blockchain that hold the potentials to enhance TRMS.

\subsection{Trust and Reputation Management Systems for CPS}
\label{sub:trms}
    In general, CPS applications involve a group of agents collecting data from physical environments and performing specific tasks based on the collected data, which includes interactions with other agents in the network. While we can assume that the majority of agents are honest, some agents may behave opportunistically to maximise their gains through dishonest behaviour. In addition, the collected data may also be noisy, faulty or maliciously tampered. Ideally, an agent should not blindly trust other agents due to these risks that may degrade the quality of service of their interaction. TRMS are designed to quantitatively assess the trustworthiness of a particular agent or data in a system through numerical and tangible values. In CPS, TRMS acts as an intermediary between service providers and requesters by providing protocols that guarantee trustworthiness in each interaction by means of authentication, resource management and access control.
    We discuss the general properties of a generic TRMS for CPS in this subsection.

    \subsubsection{Trust Derivation and Application}
    \label{subsub:trms-steps}
\begin{figure}[t]
    \centering
    \includegraphics[width=0.8\textwidth]{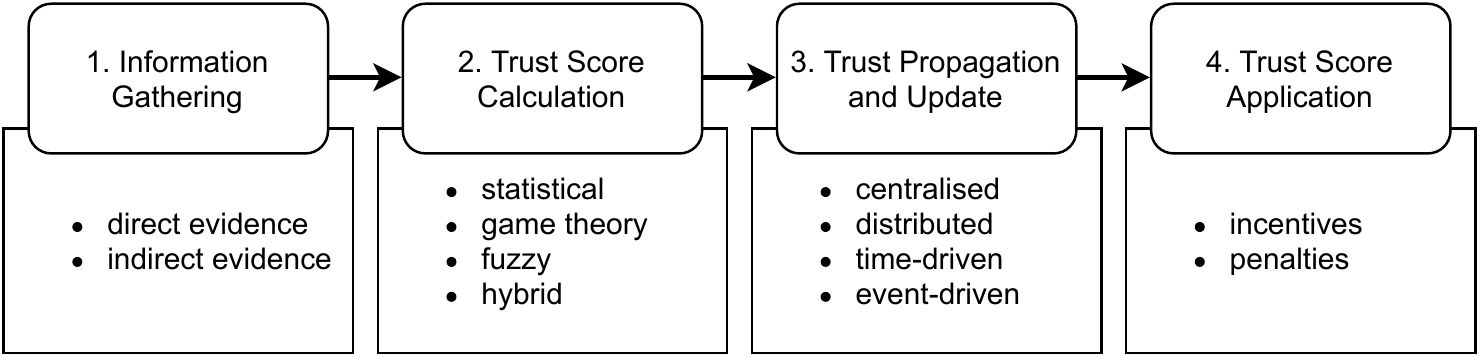}
    \caption{The steps for trust derivation in a TRMS.}
    \label{fig:trms-steps}
    \end{figure}

    Similar to real life social interactions, computational trust is built gradually from successive interactions between entities that correspond to positive or negative experiences affecting the overall belief of the trustworthiness level. In a generic TRMS, the interactions are assessed empirically, which includes four steps for collecting and applying trust computation, depicted in Fig.~\ref{fig:trms-steps}~\cite{Sharma2020}.
    
    \begin{description}
        \item[Information gathering] The first step in TRMS is defining the input parameters and attributes for quantifying or computing the trustworthiness level, which in general is highly application specific. Some of the examples include adherence to communication protocol, quality of service and degree of satisfaction towards a service. The TRMS should gather all of these input values either by 1) direct observations or interactions, or 2) recommendation from other entities if prior interactions are unavailable.
        
        \item[Trust Score Calculation] The next step includes the actual calculation of the trustworthiness level as quantifiable values or scores according to the preferred trust or reputation model. The TRMS may use various computation models that suit the application requirements, for instance, statistical, game theory, fuzzy computation or hybrid. Note that, the input attributes also determine the appropriate computation model, e.g., sum and mean models are suitable for continuous input values, while Bayesian model is more suited for discrete binary values~\cite{Sharma2020}. In addition, trust score calculation should also take into account the types of trust (see Section~\ref{subsub:trust-type}).
        
        \item[Trust Propagation and Update] Typically, trust propagation can be performed in a centralised, distributed or semi-distributed fashion, depending on the underlying architecture of the system. The trust computation should be initiated based on temporal dynamics depending on the specific application, which includes time-driven and event-driven approaches. In the time-driven approach, the trust score is updated on a regular basis, while the even-driven approach only requires updating the trust values upon new interactions and events.    
        
        \item[Trust Score Application] The specific manner in which the trust score is used depends on the requirements and operation of the application. Generally the trust score is employed to give certain quantified and fair measures for providing incentives or enforcing penalties, which may include certain privileges and monetary incentives or some restrictions and punishments. Section~\ref{sec:use-cases} discusses specific examples of how the trust scores are manifested in various CPS applications.
    \end{description}

    \subsubsection{Types of Trust}
    \label{subsub:trust-type}
    As discussed earlier, CPS applications rely on the data collected, processed and transferred in the system, and the interactions among entities. Thus, we can categorise the computation of trust as follows:
    
    \begin{description}
        \item[Behaviour-based Trust Computation] In behaviour-based trust computation, the trustworthiness level of an entity is derived from how the entity behaves in the system as perceived by a subject during its interaction with the entity. A subject identifies a positive behaviour if the observed entity conforms to the prescribed protocols and expectations, while negative behaviour corresponds to a deviation from protocols and expected behaviour.
    
        \item[Data-based Trust Computation] In data-based trust computation, the trust values are calculated based on the quality of data provided by an entity. For instance, in service-oriented CPS applications, trust can be derived from the quality of the data acquired from the data provider. Here, trust grows with the authenticity of the data, i.e., deliberate manipulation, noise or anomaly in the data would degrade the trust. In this type of trust, data validation plays an integral role and one approach to validate the data quality may include using correlated observations obtained from other entities in proximity.
   
        \item[Hybrid Approach] Relying on a single type of trust may not be sufficient for deriving trust in certain CPS applications, for instance, mobile crowd sourcing, wherein trustworthy agents are seen as those who provide reliable data and conform to the predetermined governance. In such scenarios, trust can be computed considering both the data-based and behaviour-based characteristics. In the hybrid approach, weightings are used to give favourable emphasis on either data or behaviour characteristics.
    \end{description}

    \subsubsection{Evidence Aggregation Approach}
    \label{subsub:aggregation}
    TRMS may adopt one of the evidence aggregation approaches to accumulate trust evidence and calculate the final trust and reputation score. While there is an exhaustive list of aggregation approaches~\cite{josang2007survey}, the following approaches are among the most widely adopted:
    
    \begin{description}
        \item[Sum and Mean] The most intuitive and popular aggregation approach in TRMS is summation or average of the aggregated trust evidence~\cite{Hasan2020}. Due to its simple operation, this method can also be validated manually to provide an objective confirmation. Some weighting parameters may also be incorporated to give more weight to recent or more important evidence. One of the challenges with this approach is the determination of appropriate weights, which would have an impact on the performance of the TRMS.
    
        \item[Flow Network] This approach is proposed in Advogato~\cite{levien1998attack}, wherein each participant is seen as a node in the network, while the interactions between participants are modelled as network flows. Consequently, the trust is derived from the number of flows a participant obtained from others. This approach is relatively robust to trust-related attacks, as the total active flows in the network are assumed to be constant and strictly regulated by the TRMS.
    
        \item[Markov Chain] As implemented in EigenTrust~\cite{kamvar2003eigentrust}, Markov chain approach works based on probability modelling of a user's feedback reaching at a particular participant. The feedback from one user to another is modelled as a probability function of a transition from source to target user, using which the reputation score is derived from.
    
        \item[Bayesian] In this approach, the trust and reputation score are computed using statistics. The trustworthiness score is described as a beta distribution of two parameters where $\alpha$ and $\beta$ denote positive and negative recommendation, respectively. To calculate and update the score, an update to the provided beta distribution is performed, through which unfair ratings can also be removed~\cite{whitby2004filtering}.
    \end{description}

    \subsubsection{Trust Dimensions}
    \label{subsub:dimension}
    In general, trust is strongly attached to a particular context and generally cannot be transferred to another context without rigorous adjustment and re-calculation. With this regard, context-awareness is an important factor to consider in designing a TRMS. A TRMS can work on a single context or multiple-context awareness in deriving trust from collected evidence depending on its initial design~\cite{Bellini2020}.
    
    \begin{description}
        \item[Single-dimension] A lightweight TRMS with a simplified trust and reputation model might only incorporate a single-dimension trust evaluation for the sake of limited resources in CPS applications. While single-dimension trust model may not be comprehensive, it may be preferred depending on the application design, e.g., when a majority of resource-constrained devices are in use.
    
        \item[Multi-dimension] On the other hand, multi-dimensional trust and reputation model represents trust and reputation scores in multiple parameters or a single value derived from multiple parameters with appropriate weightings. In practice, this type of TRMS may require heavier computation and may not be suited for constrained devices.    
    \end{description}

\subsection{Adopting Blockchain for TRMS}
\label{sub:blockchain-for-trms}
Since its initial inception in 2009 as a pioneer in decentralised cryptocurrency, blockchain has also been applied in many non-monetary applications, one of which includes TRMS. While blockchain may introduce some overheads, blockchain has promising potentials to be implemented in TRMS. Here, we describe the inherent properties of blockchain that would bring enhancements and benefits to TRMS.

\begin{figure}[t]
\centering
\includegraphics[width=0.8\textwidth]{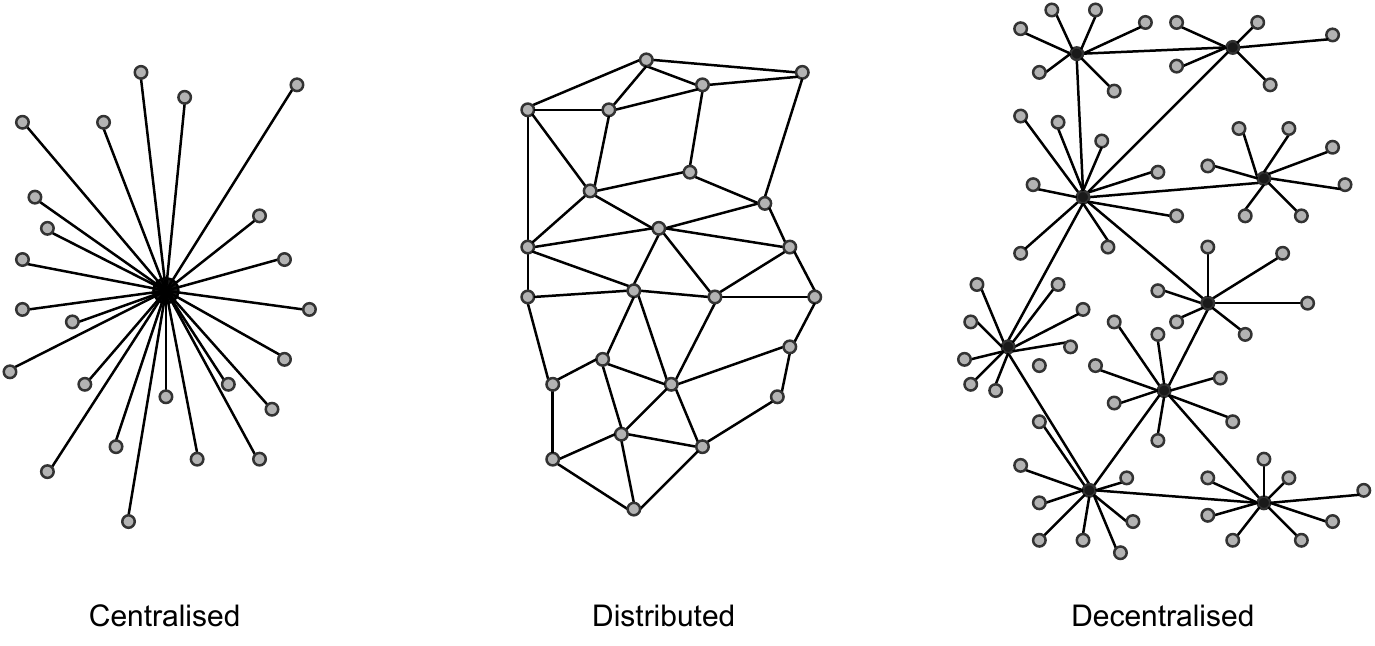}
\caption{Different types of architecture: centralised, distributed and decentralised.}
\label{fig:types-architecture}
\end{figure}
    
    \subsubsection{Decentralisation}
    \label{subsub:decentralisation}
    Conventional TRMS relies on a third party aggregator to collect trust evidence and calculate trust scores. Trusting a third party aggregator actually introduces significant risks, for instance, when the aggregator is compromised any underlying processes of trust computation could be maliciously altered and the sensitive data could be in danger. On the other hand, blockchain removes any Trusted Third Party (TTP) and comes with a decentralised architecture, as seen in Figure~\ref{fig:types-architecture}, which eliminates associated risks of employing third party aggregators in TRMS. Blockchain employs certain consensus algorithms, such as Proof-of-Work and Proof-of-Stake, to enforce collaborative execution and validation of transactions, using which fraudulent manipulations can be avoided. The ledgers, in which the transactions or data are stored, are replicated among all participants in the network that enhances availability. In addition, blockchain can also be incorporated in a distributed TRMS to enhance the mechanism, for instance, by utilising smart contracts, which we describe in the following subsections.
    
    \subsubsection{Smart Contract}
    \label{subsub:smart-contract}
    Smart contract is a form of execution codes agreed by a set of users, which allows deterministic and trusted execution of business logic with reliable guarantees that the process would be accomplished and validated collaboratively in the network. Smart contracts can be embedded into a blockchain-based TRMS to perform collection and calculation of trust scores which can offload the trust computation from the CPS devices. For instance, a node may submit feedback to the smart contract about the experience interacting with a service provider, which later will be used to calculate the service provider's reputation score. Another node in the network can also query the smart contract to obtain the reputation scores of particular service providers. That is, a smart contract acts as a reliable intermediary for computing and querying trust related information.
    
    \subsubsection{Pseudonyms}
    \label{subsub:pseudonyms}
    There is an inherent risk of leaking sensitive information in conventional TRMS, as the authentication mechanism may link the identification details to real-life identities. Blockchain introduces an elliptic curve public key cryptography mechanism which utilises pseudonyms, i.e., public key, for identification purposes, resulting in higher privacy preservation, as real identities are not used. The use of pseudonyms is, to some extent, beneficial for protecting users' privacy, which is a desirable property in designing a TRMS. In a blockchain-based TRMS, each node is identifiable by its public key which hides any personal data, such as device ownership details. We discuss more challenges and opportunities in privacy preservation for DTRMS in Section~\ref{sub:challenges-privacy}.
    
    \subsubsection{Immutable Storage}
    \label{subsub:immutable-storage}
\begin{figure}[t]
\centering
\includegraphics[width=0.85\textwidth]{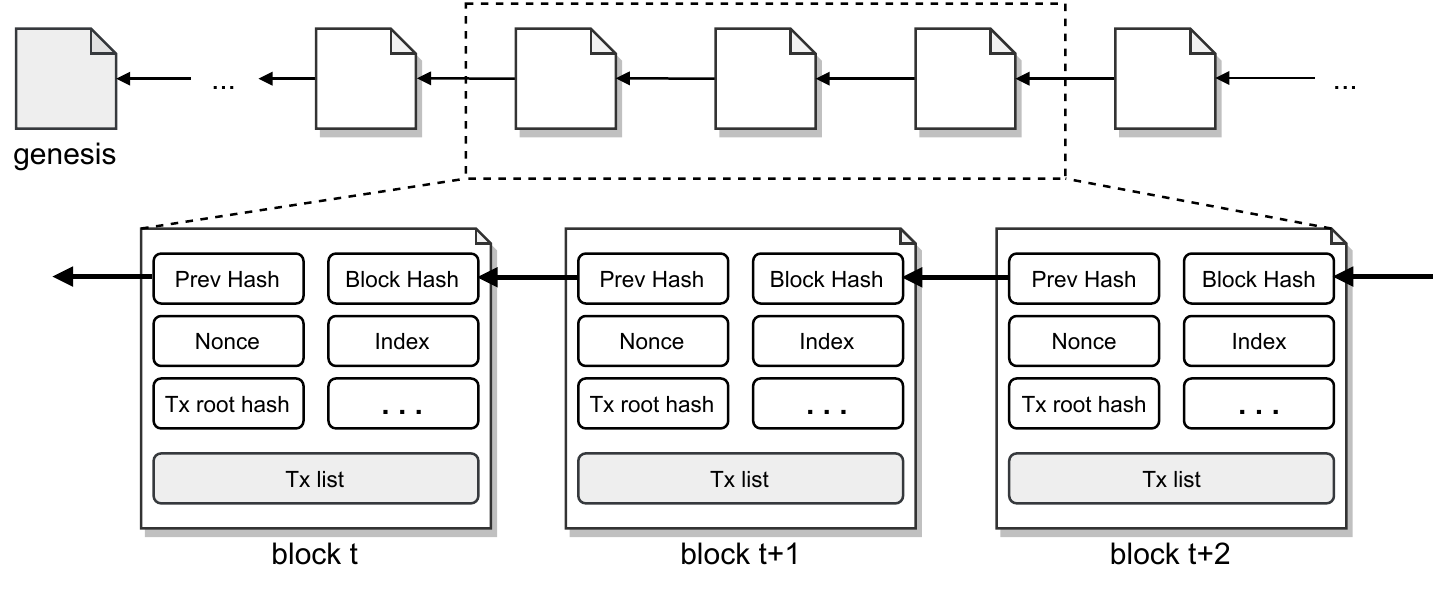}
\caption{An overview of blockchain's immutable storage.}
\label{fig:immutable-storage}
\end{figure}

    Traditional TRMS stores trust evidence and interaction history on each device's internal memory, which may overwhelm the devices, especially if there is a large amount of information in a network with thousands of nodes. As discussed earlier, a traditional TRMS can also rely on a TTP to keep track of the trust related information, but with fundamental risk of data loss and manipulation linked to the centralised approach. Blockchain data structure, as depicted in Figure~\ref{fig:immutable-storage}, enforces immutability as it is difficult if not almost impossible to tamper the stored data on the blockchain. To tamper the data, an attacker should break the hash cryptography and may need to traverse all the way back to the genesis block. With proper removal of any Personally Identifiable Information (PII), blockchain is a perfect and safe solution to store interaction evidence that would later be used to calculate the trust score.
    
    \subsubsection{Transparency}
    \label{subsub:transparent-mechanism}
    Recall that in conventional centralised TRMS, any underlying mechanism in trust and reputation calculation is performed by a centralised aggregator, which conceals the actual process from other participants in the network. On the other hand, blockchain offers transparent mechanisms in collaborative trusted execution of business logic via smart contracts and transparent immutable storage via a transparent shared ledger. With precautions in handling and storing sensitive information in the ledger, this type of transparent mechanism is preferred as it enables a traceable source of evidence where any participant can ascertain the integrity of a trust calculation by examining the ledger.

\section{Use Cases}
\label{sec:use-cases}
In this section, we outline various implementations of DTRMS for CPS across different application domains to demonstrate how blockchain enhances and brings benefits to these applications. We also present a summary of the use cases in Table~\ref{tab:dtrms-use-case-summary}.

\begin{table}
\caption{Summary of DTRMS use cases.}
\label{tab:dtrms-use-case-summary}
\begin{tabular}{ p{4cm}  p{7cm}  p{3cm} }
\hline\noalign{\smallskip}
\textbf{Application Fields} & \textbf{Goals} & \textbf{Blockchain Features}  \\
\noalign{\smallskip}\hline\noalign{\smallskip}
Generic CPS & 
    end-to-end IoT trust establishment; securing CPS/IoT & 
    \quad IS, AC, PS, SC \\ 
    \noalign{\smallskip}
Supply Chain Managements & 
    managing traders and sellers; maintaining product quality & 
    \quad SC, IS, PS \\
    \noalign{\smallskip}
Crowd Sourcing & 
    selecting reliable workers &
    \quad AC, IS, PS \\
    \noalign{\smallskip}
Robotic and Autonomous Systems & 
    selecting reliable service providers and detecting Byzantine nodes & 
    \quad SC, IS, TR \\
    \noalign{\smallskip}
Vehicular Ad-hoc Network & 
    validating exchanged messages to avoid malicious messages & 
    \quad PS, AC, IS \\
    \noalign{\smallskip}
IoT Data Marketplace & 
    curating traded data and providing fair payments & 
    \quad SC, PS, IS \\
    \noalign{\smallskip}
Distributed Energy Trading & 
    managing prosumers' reliability & 
    \quad SC, AC, IS \\
\noalign{\smallskip}\hline\noalign{\smallskip}
\end{tabular}
IS = Immutable Storage, AC = Adaptive Consensus, SC = Smart Contracts, PS = Pseudonyms.
\end{table}

\subsection{Generic CPS Trust Architecture}
\label{sub:generic}
Blockchain ensures trusted, persistent and immutable storage for keeping observational data which achieves tamper-proof storage and prevents unwanted or malicious modification. However, the fundamental problem about establishing trust in the data itself cannot be solved only by using blockchain. In~\cite{Dedeoglu2019}, the authors propose an end-to-end trust architecture for IoT. The authors also proposed a dynamic block validation mechanism, wherein trust management reduces the computation load on the nodes by reducing the amount of transactions that need to be validated from trusted nodes.

In~\cite{Yan2021}, the authors proposed a blockchain-based trust evaluation system for Pervasive Social Networking (PSN), which helps to protect a node from unfamiliar or unknown acquaintances in a trustless decentralised environment. In the proposed solution, the trustworthiness of a participant is derived from its social behaviour, represented as trust evidence stored on the blockchain.  The authors also proposed a trust-based consensus algorithm that aims to reduce resource consumption and accelerate block generation in the consensus process. A new block is confirmed if it is approved by an adequate number of miners above a certain threshold.

In~\cite{kouicem}, a layered architecture, called BC-Trust, is proposed to provide a scalable DTRMS solution for devices with high mobility in fog computing. In BC-Trust, the trustworthiness of a particular service provider is derived from a user's direct experience and recommendations from known peers in the network. Fog nodes with sufficient computing resources maintain the blockchain and perform all required trust computation, offloading the trust computation to high resource devices, which help to reduce the load in the constrained devices. BC-Trust is also designed to be robust against known trust-related attacks, such as ballot-stuffing and bad-mouthing (see Section~\ref{sub:challenges-security}).

A trust architecture can also be utilised to protect important CPS resources from illegitimate access by unauthorised service consumers. In~\cite{putra-icbc2020}, the authors proposed a decentralised Attribute-based Access Control (ABAC), in which each service consumer is associated with a certain trust score based on its behaviour in the network. The trust score of the service consumer is then included in the required attributes to access resources. The authors introduced three smart contracts, namely attribute provider, trust and reputation system and policy smart contract to administer and operate the trust-based access control. In~\cite{putra-tnsm}, the authors extended their solution to include privacy preservation and more efficient trust computation, where the computation is offloaded to the blockchain entirely.

\subsection{Supply Chain Management}
\label{sub:supply-chain}
In general, Supply Chain Management (SCM) systems demand for traceability, which is fulfilled by the blockchain. In addition, blockchain allows multiple writer to the system which is suitable in most instantiations of supply chain wheres multiple stakeholders are involved. DTRMS is used in supply chains to provide additional trustworthiness and benefits, as the quantified trust score could be associated with certain product qualities or producers.

Bai et al.~\cite{Bai2021} proposed a trust management scheme in an e-agriculture supply chain scenario where a network of smart greenhouses act as miners and form the blockchain network. Each greenhouse manages a set of sensors that monitor the condition of the farm. The farmers can use the network to query the sensors to get some agriculture related information, such as the probability of whether the farm needs to be fertilised or watered. The authors proposed a DTRMS in which a game theoretic approach is used to determine the trustworthiness of each sensor's reading by associating a trust score to each sensor. When a farmer queries for sensor data, the green house will also look for other readings from related sensors and perform Bayesian inference to report an estimate along with a trust score. A low trust score indicates that the readings are incorrect and the sensor may need to be replaced.

Although blockchain can solve the immutability and traceability issues in supply chain applications, the issue of the integrity of inserted data remains unsolved. In~\cite{sidra-trustchain}, the authors proposed TrustChain, a three layered trust management framework tailored for supply chains. The architecture consists of three layers, namely data layer (for data input), blockchain layer (where all process happens), and application layer (for transacting with blockchain). In general, the reputation system assesses the quality of the commodities based on multiple observations within the supply chain. The solution adopts smart contracts to automate reputation calculation and specifically deploys two contracts: 1) a quality contract to assess the quality of each supply chain commodity based on sensor readings (e.g. temperature to keep the food quality) and 2) a rating contract to compute the reputation of the traders. Each trader has an inherent trust score, which is derived from all ratings with customisable weightings. The method also utilises time-varying and amnesic trust calculation where more emphasis is given to recent observations.

In~\cite{Li2020}, the authors proposed Reputation-based Trustworthy Blockchain Supply Chain Management (RTB-SCM) to address trustworthiness issues in supply chains. The solution is based on a consortium blockchain, named Reputation Assessment Blockchain (RAB), that stores trade records and commodity information. RAB introduces a token based reputation system, which is based on a cryptoasset governed by the trusted regulator that runs the consortium blockchain. In addition, the design utilises a smart contract based reputation rating model that utilises tokens for rewards and punishments. The authors proposed two algorithms; 1) Quality Status Generation (QSG) for quantifying $QInfo$, the quality of a trade with regard to a commodity type derived automatically by the sensors and 2) Token-based Reputation Reward/Punishment (TR2P) for determining the appropriate reward or punishment based on $QInfo$ from QSG.

\subsection{Crowd Sourcing}
\label{sub:crowd-sourcing}
In~\cite{Zou2019}, the authors proposed a hybrid blockchain architecture to enhance data validation in crowd sourcing, wherein a private consortium blockchain is used as the backbone of the network, while the public blockchain acts as a method of transaction validation for the novel consensus protocol. In this work, trust management is incorporated in a consensus algorithm called Proof-of-Trust (PoT). The intuition is to select reliable validators to validate collected data based on the trustworthiness score of the participants in the crowd sourcing service. Combined with RAFT leader election and Shamir's secret sharing algorithm, the framework calculates the trust score based on three independent parameters: 1) the number of transactions the user has on the platform, 2) the total time the user has been involved in the validation process and 3) the number of complaints the user receives. The PoT protocol splits the consensus process into four phases, each of which is conducted by different roles, which ensures performance and consistency of the consensus process while greatly improving scalability.

In ~\cite{Feng2019}, Feng and Yan proposed MCS-Chain, which is a fully decentralised trust management for Mobile Crowd Sourcing (MCS) purposes without relying on trusted actors.. The architecture consists of end users, workers and miners. The blockchain acts as the MCS platform, where all the procedures are recorded and trust scores are evaluated. The miners in this case are the cell towers which are responsible for managing the blockchain. A trust evaluation scheme is designed to help the end users choose appropriate workers based on reliability. End users post some particular tasks to the blockchain by invoking blockchain transactions, which then broadcast the task to the workers for bidding. In this step, the trust score helps the end users to pick the preferred workers. The trust score of the workers are then updated based on the feedback from the end users about certain tasks. To avoid unfair rating, the mechanism applies deviation between personal and average feedback and also considers the previous trust score of the submitter. The authors implemented their solution on Android and Windows to evaluate the performance and highlight the efficiency of the proposed system.

\subsection{Robotic and Autonomous Systems}
\label{sub:robotics}
DTRMS has also been implemented in the area of robotics and autonomous systems, where blockchain overcomes several reliability issues in information sharing and aids the selection of service providers in a transparent way.

In~\cite{alowayed2018}, Alowayed et al. proposed a custom DTRMS which enables Autonomous Systems to evaluate network providers based on their ability to provide reliable interconnection service as per the pre-approved Service Level Agreement (SLA). In this framework, the network performance measurements are stored as transactions on a permissioned blockchain. A smart contract then quantifies the trustworthiness of each network provider, called SLA score, and analyses if the provider has submitted a misleading performance report. They propose to use the SLA scores to select network providers to ensure that clients are provided the required quality of service. The framework requires each SLA score to be written on the blockchain with a transparent and publicly-agreed SLA score calculation method between network participants, while privacy preservation is achieved by adopting order-preserving encryption mechanism ~\cite{boldyreva2009order}.

Strobel and Dorigo proposed a blockchain-based knowledge sharing architecture designed for swarm robotics~\cite{dorigo2018blockchain}. In this framework, a DTRMS is employed to identify Byzantine or malicious robots which may hinder the overall performance of the system due to misleading data measurements. A permissioned blockchain network is utilised, wherein each robot serves as an Ethereum node, through which the robot could exchange knowledge to other robots within 50 cm of proximity via blockchain transactions. The reputation for each robot is calculated based on the absolute difference between reported observation and the average of other observations from other robots in the proximity.

\subsection{Vehicular Ad-Hoc Networks}
\label{sub:vanet}
Several works have been proposed in the field of Vehicular Ad-Hoc Network (VANET) to incorporate DTRMS for enhancing the security or avoiding malicious events in the network. The typical architecture of VANET includes mobile and fixed nodes, e.g., the smart vehicles and Road Side Units (RSU), respectively.

The authors of~\cite{Zhao2021} proposed a privacy preserving announcement protocol for Internet of Vehicles (IoV) called PBTM. The authors also designed a blockchain-based trust management system to ascertain the authenticity and synchronise the timing of the exchanged messages. RSUs play an important role of maintaining the blockchain and calculating the trustworthiness score of each vehicle using weighted sums method according to the validity of the transmitted message. Privacy preservation is achieved by adopting an identity-based group signature, which realises the anonymity of each vehicle.

Wang et al. proposed BSIS: Blockchain-based Secure Incentive Scheme, a reputation-based consensus protocol to obtain efficient consensus within a Vehicular Energy Network (VEN)~\cite{Wang2019}. In the proposed consensus algorithm, the validators are selected based on their trust score and each validator would receive incentives upon successful execution of the consensus mechanism. Consequently, the higher the trust score, the more chance a node has to be selected as a validator and subsequently obtain the reward. The authors classify two types of trustworthiness scores, namely local trust and reputation value. The local trust score is derived from ratings obtained from each interaction with other energy nodes, while the final reputation value is calculated by aggregating all local trust scores that a node has obtained.

In~\cite{Yang2019}, the authors proposed a blockchain based trust management for VANET, wherein RSUs are the only approved actors to compute the trust values for each vehicle. In this framework, each vehicle may send messages to other vehicles from which the trust score of the vehicle is calculated. Each message receiver generates ratings that represents the credibility of the corresponding message. Due to high mobility and limited storage capacity, each vehicle is expected to periodically submit the ratings to nearby RSUs, in which ratings are aggregated and grouped to calculate the trust score of each vehicle using Bayesian inference. Once the score is stored on the blockchain, any vehicle may query the data if necessary. The authors argue that the proposed DTRMS would give guidance to the vehicles about the quality of the received messages and also provide the underlying evidence for reward and punishment mechanisms.

Lu et al. proposed a trust management system for VANETs, called BARS: a Blockchain-based Anonymous Reputation System~\cite{Lu2018}, with an emphasis on anonymity by avoiding linkability between real and pseudo identities. In this framework, the trustworthiness of each vehicle is determined from the authenticity of the broadcast message and the reported opinion from other vehicles. The authors introduced the concept of Law Enforcement Authority (LEA) which is responsible for managing the framework and resolving disputes.

\subsection{IoT Data Marketplace}
\label{sub:iot-marketplace}
The proliferation of CPS/IoT deployments has generated enormous amount of data, using which the owner can obtain financial benefits by selling useful data to specific consumers. In an IoT data marketplace, both sellers and consumers can communicate and share data. However, the data customers do not trust the sellers as the data quality cannot be guaranteed. Blockchain in this case can enhance IoT data marketplace by providing a decentralised payment mechanism using built-in cryptocurrency and utilising DTRMS to assess the trustworthiness of each participant.

Camilo et al. proposed a blockchain-based data trading platform, in which trust and reputation play an important role in helping customers determining the quality of the sellers~\cite{camilo2020}. The authors define a distinction between trust and reputation, where trust corresponds to buyer's view of a seller based on his trading experience, while reputation is an aggregated view of a particular seller from multiple individual trust scores across different buyers. In this platform, the data owners or sellers may advertise the metadata via a smart contract on the blockchain, which can be be explored by data buyers to select the preferred sellers. The data buyers are required to submit a feedback that rates the seller and the data quality through a feedback transaction to the blockchain, using which the smart contract evaluates the trust and reputation score of the sellers accordingly.

In~\cite{javaid2020}, the authors proposed an automatic review system to assess the quality of the data, which is used for monetisation of IoT data. The system adopts a publish-subscribe mechanism, where a MQTT broker plays an important role along with the permissioned blockchain. A rating is associated to each data on each topic in the platform. Any data buyer may request the smart contract to browse available data based on subscribed topics, with reviews associated with each data sale. Upon completing the data access, each buyer is required to submit reviews about the accessed data for an incentive. The system utilises a smart contract to achieve an automated payment and incentive mechanisms that eliminate the need for a TTP, while also reducing the associated risks of trusting an external party.

The authors in~\cite{sonbol-icbc2021} proposed a reputation system for online market places, which is based on hashcash proof-of-work algorithm, originally designed to reduce spams in email~\cite{back2002hashcash}. The reputation system is designed to assess and incentivise watchtowers to behave rationally. A watchtower is an independent entity that preserves the client's interest for specific purposes, e.g., lightning payment network~\cite{poon2016bitcoin}, by continuously monitoring the blockchain network on behalf of the clients that may frequently be offline. The watchtowers offer a monitoring service in the open market, where clients would tend to pick the watchtowers with the highest reputation score. During the negotiation phase with a client, a watchtower publishes a smart contract as a persistent proof that bonds the terms and conditions of payments and services with the client. The reputation of the watchtowers is derived from successful transactions, while a proof-of-breach is generated from the contract if a watchtower does not fulfil its obligation.

In~\cite{Zhao2018}, the authors proposed a payment mechanism for IoT marketplace, called Secure Pub-Sub (SPS), where blockchain is utilised to provide fair payments and reliability. In SPS, blockchain performs as a payment gateway between a publisher and subscriber that do not necessarily have to trust each other, wherein a subscriber can deposit some funds prior to subscribing to a particular publisher to access the data. In addition, a reputation system is employed so that each subscriber can assess the publisher after accessing its service, where a smart contract transparently maintains the process. Here, the reputation system can help the subscribers to pick appropriate publishers based on certain reputation scores that are higher than a threshold for determining reliable publishers.

\subsection{Distributed Energy Trading}
\label{sub:distributed-energy-trading}
Distributed or peer-to-peer energy trading is a marketplace where each prosumer, a user who consumes and produces energy, can transact energy to the end user directly without the need of a central entity. It has been demonstrated by recent works that TRMS can help improve the efficiency and enhance the fairness of energy trading.

In~\cite{Yahaya2020}, the authors proposed a secure blockchain-based energy trading platform with a built-in reputation system to enhance the reliability and encourage honest behaviour among blockchain nodes. The authors also incorporate  reputation scores into the Proof-of-Work consensus algorithm, called PoWR, to reduce the block creation time and overall latency. Each participant utilises the blockchain network as a communication channel, through which each participant exchanges information about direct and indirect trust experiences. In this reputation framework, a higher reputation score corresponds to higher probability of participating in the PoWR, hence more chance of getting incentives.

Khorasany et al. proposed a peer-to-peer framework for energy trading in~\cite{Khorasany2021}, where blockchain and smart contracts are employed to build a decentralised trading mechanism. The authors also associate a reputation factor to each energy agent that represents its reliability in fulfilling the obligations. In addition, the authors also introduced an algorithm, called Anonymous Proof of Location (A-PoL), to anonymously prove user's location. In performing the energy trading, each user may select a partner based on their preference on both reputation factor and agent's location, which is handled by an automated algorithm. A Dispute Resolution smart contract is in charge of calculating each agent's reputation factor which is based on prior commitments in delivering energy to the trading counterpart.

A framework called Reputation for Blockchain-based energy Trading (RBT) is proposed in ~\cite{Wang2021}. RBT utilises blockchain as a traceable and immutable storage for reputation scores and smart contract for automated reputation calculation. Here, the reputation is derived from the behaviour of each node according to its role in the P2P process via three parameters, namely role, rule and reputation. A matchmaking strategy based on $k$-double auction algorithm is used to connect both buyers and sellers and to decide trading prices that are more beneficial to both parties. The matchmaking strategy also includes a fairness indicator, which is a ratio between reputation score and the average income and cost for sellers and buyers, respectively.

\section{Challenges and Future Directions}
\label{sec:challenges}
There are many open research directions for DTRMS. In this section, we discuss several issues and challenges that need to be addressed for future research on DTRMS, which include scalability, privacy, excessive resource consumption, security and interoperability.

\subsection{Scalability}
\label{sub:challenges-scalability}
Blockchain requires a transparent shared ledger which is replicated between blockchain participants to enforce redundancy and maintain consistency. While the replication removes a single point of failure and increases availability, the shared ledger may grow significantly due to the append-only nature of blockchain storage mechanism. In the long run, the explosion of storage requirement may hinder the performance of the network as it demands high memory requirements, which causes high synchronisation times for new node initialisation. For instance, as of June 2021, Ethereum blockchain size has reached approximately 820GB for a default full node\footnote{\url{https://etherscan.io/chartsync/chaindefault}} and 7.5 TB for a full archival node\footnote{\url{https://etherscan.io/chartsync/chainarchive}} and is expected to grow by approximately 75GB per year.

In addition, blockchain is known to have a limitation in block generation time, which contributes to the scalability issues. While Visa payment network could achieve up to 47,000 transactions per second (tps), Bitcoin is only able to cater approximately 7 tps with limited block size of 1 megabyte~\cite{poon2016bitcoin}. The fixed rate of block generation time introduces a bottleneck, and could be exacerbated if there are more transactions to be processed.

DTRMS is thus faced with potential scalability issues inherent to the underlying blockchain. The growing number of participants in the network would also deteriorate the scalability of DTRMS, as accommodating a large number of nodes results in large storage requirements. It is known that addressing scalability for blockchain is still an open research problem and the community is still actively proposing new solutions. Some methods to overcome scalability issues in a DTRMS may include:

\begin{figure}[t]
\centering
\includegraphics[width=0.8\textwidth]{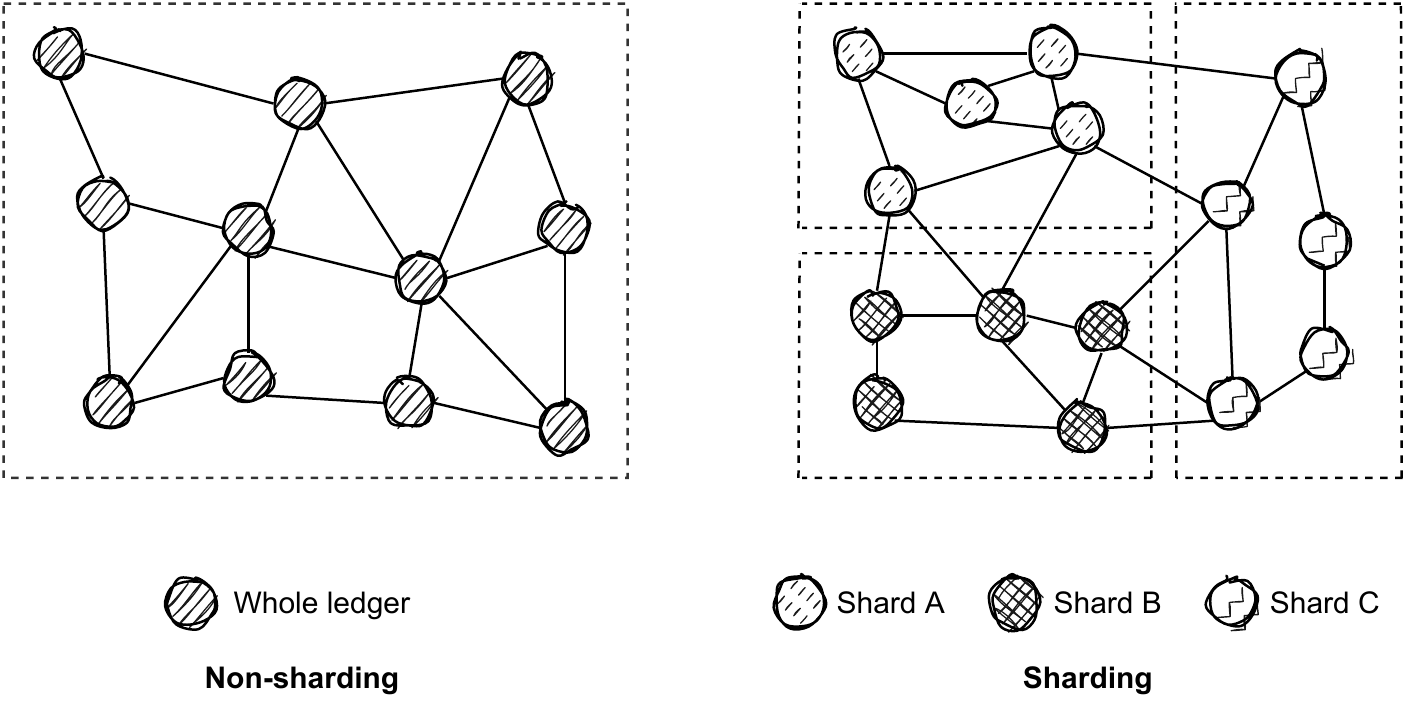}
\caption{The intuition of sharding for improving scalability.}
\label{fig:sharding}
\end{figure}

\begin{description}
    \item[Sharding] In this mechanism, a full copy of the shared ledger that represents the current state of the blockchain is separated into several chunks which are distributed to different nodes. That is, each node would retain different parts of the whole ledger which significantly reduces the storage requirement to store the entire blockchain state (see Figure~\ref{fig:sharding}).

    \item[Scalable Reputation Scheme] The reputation model can be adjusted to circumvent the scalability issue by storing a collective or aggregated trust information that resembles a group of devices instead of storing trust information for all devices, which may be redundant~\cite{Sharma2020}. These storage mechanisms may help reducing the amount of data stored in the ledger, which helps to alleviate the scalability issue. For example, the reputation model may average the trust score of a group of devices from the same owner prior to storing  the record in the ledger.
    
    \item[Off-chain computations] As demonstrated in the Lightning network of Bitcoin~\cite{poon2016bitcoin}, off-chain computations would result in significant reduction of processing latency, as the transactions are processed off-chain without going through excessive consensus processes. In this scenario, two parties retain a signed contract that resembles the current token balance of each party and continue the transactions off-chain by keeping a signed log of balance transfers. An overlay network can also be constructed from multiple-signed contract of different owners, i.e., the so called Lightning network. The contract is ceased when a party submits a final transaction to the main ledger that transfers the tokens according to what is recorded in the signed logs.
\end{description}

\subsection{Privacy}
\label{sub:challenges-privacy}
One of the main motivations of incorporating blockchain for TRMS is the inherent use of pseudonyms instead of real identities for authentication purposes. Using public keys as pseudonyms for authentication would conceal the actual identities of the users and make it difficult to link the public keys to the real user identities, a preferred requirement for privacy preservation. However, research has shown that it is possible to track user behaviour from pseudonym-based transaction logs and link them back to the real identities~\cite{Clemence2019}. Ideally, a DTRMS should achieve privacy preservation by addressing the following concerns~\cite{Hasan2020}:

\begin{description}
    \item[User anonymity] The goal of user anonymity is to conceal the actual identity and prevent linkage attacks. For example, a user can be represented by more than one pseudonym, e.g., replaceable public keys, that would allow the users to continue transacting in the system without allowing a malicious entity to link the multiple pseudonyms to their actual identities.

    \item[Feedback confidentiality] In practice, total user anonymity is difficult to achieve and some information regarding the identities or interactions of users might be inferred from the data revealed by the applications. However, it is a critical requirement to ensure the confidentiality of feedback information submitted by the users in order to encourage users to submit truthful feedback.
\end{description}

The research community in the field of security and privacy still actively proposes new methods for privacy preservation, some of which can be implemented for DTRMS:

\begin{description}
    \item[Zero-Knowledge Proofs] Using Zero-Knowledge Proofs (ZKP), one can verify the validity of a statement by getting a plausible proof without exposing any additional information other than the validity of a statement, hence preserving the privacy~\cite{goldwasser1989knowledge}. ZKP can be implemented to provide a means to validate a particular pseudonym without any risk of revealing sensitive information that might be linked back to the real identity of a user.

    \item[Homomorphic Encryption] Typically, a generic encryption algorithm provides a means to conceal sensitive information to preserve privacy. However, the encrypted data, i.e., ciphertext, is practically unusable and decryption is required for the ciphertext to be usable, which may expose sensitive information to public. Homomorphic encryption, on the other hand, allows computations on encrypted data without the need of decryption, which may help to keep the sensitive information hidden. For instance, in DTRMS homomorphic encryption permits several operations on encrypted feedback to achieve feedback confidentiality.
    
    \item[Secure Multi-Party Computation] Hiding sensitive information can be achieved by utilising Secure Multi-Party Computation (SMPC), using which several input values can be aggregated into an output value without revealing the individual input values. SMPC can be implemented in a DTRMS to calculate trust or reputation scores by aggregating several feedback values, while keeping the feedback values private.
\end{description}

\subsection{Resource Consumption}
\label{sub:challenges-resource-consumption}
The traditional centralised TRMS approach utilises a single TTP that acts as a single authoritative entity for managing resources and making decisions. While there is an inherent risk in trusting a TTP, this centralised architecture results in a relatively low resource consumption. On the other hand, the absence of TTPs in DTRMS requires a distributed consensus algorithm that typically consumes high resources while also sacrificing latency and throughput. For instance, Bitcoin is notorious for its low block generation rate of 10 minutes and high carbon footprint from the coin mining process~\cite{stoll2019carbon}.

In general, CPS applications may consist of thousands of interconnected constrained devices that demand high throughput and low latency. One possible solution to overcome the high resource consumption of blockchain is to devise a tailored consensus algorithm for CPS that suits the constraints of typical CPS devices~\cite{dorri2016blockchain}. In~\cite{dorri2019lsb}, the authors proposed a blockchain platform tailored for CPS/IoT architecture, called Lightweight Scalable Blockchain which (LSB), where the computation and storage capacity are typically constrained. LSB employs a lightweight consensus algorithm and distributed throughput management specifically designed for CPS/IoT.

As discussed in previous sections, a reputation system can be implemented to create an efficient consensus algorithm with less overheads. In reputation-based consensus algorithm, a Proof-of-Work consensus algorithm that demands solving a mathematical puzzle for verifying a block is replaced with a Proof-of-Stake like consensus where the block validator is selected based on its reputation in the network~\cite{repucoin}. While reputation-based approach is still prone to trust or reputation attacks (see Section~\ref{sub:challenges-security}), it can reduce the computational load for achieving a distributed consensus to determine the next block to be mined.

\subsection{Security}
\label{sub:challenges-security}
As both traditional and decentralised TRMS rely on multiple untrusted parties to gain collective knowledge for building reputation scores, any entity in the system can launch attacks that may impact the normal operation of the system. There are known attacks in both traditional and decentralised TRMS, which are commonly referred to as trust-related attacks. A DTRMS should be resilient to these attacks, as noted in the following:

\begin{description}
    \item[Sybil Attack] In this type of attack, an adversary creates several forged identities that can be utilised to gain disproportionate influence against a benign user with a single truthful identity. The adversary may illegitimately utilise the forged identities to launch attacks such as ballot-stuffing or bad-mouthing for its own benefits. Sybil attacks can be avoided by increasing the cost of creating new identities~\cite{douceur2002sybil}. Also, when privacy is not a key requirement, linking to real identity is proven to be effective in preventing Sybil attack~\cite{yu2008sybillimit}.

    \item[Ballot-stuffing] An adversary can illegitimately increase its reputation score by launching a Sybil attack to submit multiple fake feedback or by colluding with other adversaries. This attack is called ballot-stuffing and is also often referred to as self-promotion attack. The risk of ballot-stuffing attack can be partly reduced by using coins to submit feedback~\cite{Bellini2020}, which would impose a significant cost to the adversaries for submitting multiple feedback by themselves, while the advantage from the attack may not be worthwhile.
    
    \item[Bad-mouthing] In contrast to ballot-stuffing, bad-mouthing is an attack that aims to ruin another honest user's reputation score by providing negative feedback regardless of the behaviour of the target user. Bad-mouthing attacks could lead to severe damaging effects, especially for sensitive applications, such as monetary systems~\cite{Hasan2020}. While mitigating this effect is a non-trivial task, one possible protection could be to compare feedback from unknown entities to those from highly trusted nodes~\cite{josang2009challenges}.
    
    \item[Whitewashing] When the adversaries have low or negative reputation scores, they can rejoin the system with a new identity resulting in a fresh reputation score. This type of attack is attractive to the adversaries especially when the cost of re-entering the system is very minimal. As a mitigation scheme, the system may require users to link the identity or pseudonym with a real-world identity, e.g., a website, that would incur significant cost for modification~\cite{hoffman2009survey}.
    
    \item[On-off Attack] The adversary may act opportunistically by providing alternating feedback, i.e., positive and negative, to maintain the reputation score at a safe level to avoid detection. For instance, the adversary may constantly provide positive service to get selected as a service provider, but once selected, the adversary launches the attack by providing poor services to the selected target nodes~\cite{Chahal2020}. Appropriate weighting according to temporal dynamics in the reputation formula would help to reduce the risk of this attack~\cite{Hasan2020}. In this approach, higher weights are applied to recent or more important interaction evidence, resulting in significant decline of the trust score in case of an attack.
\end{description}

\subsection{Interoperability}
\label{sub:interoperability}
In practice, it is common for CPS applications to have a wide range of technological implementations with different types of hardware and protocols for communications. Consequently, these implementations tend to work in isolation with very minimal cross-platform collaboration. Interoperability is an important factor to consider in designing CPS solutions for achieving efficient collaboration across platforms and applications. For instance, in a blockchain-based smart city architecture, interoperability would allow transfer of digital assets and values across different blockchain platforms~\cite{Dedeoglu2020}. There are several initiatives in blockchain interoperability, which include Cosmos\footnote{\url{https://cosmos.network/}} and Interledger\footnote{\url{https://interledger.org/}}.

In addition, carefully modelled interoperability schemes would also allow transfer of trust and reputation scores across different platforms. In any TRMS application, the lack of prior interactions or recommendations is a challenge in quantifying reputation, which could be mitigated by transferring reputation values from a separate platform that has already gathered some evidence about the trustworthiness of a particular entity. Note that context awareness should also be taken into account in this mechanism, as trust is highly associated with the underlying context and may not be transferable from one context to the other.

In the context of DTRMS for CPS, interoperability would also mean an ability to appropriately derive trust and reputation scores across heterogeneous constrained devices. As typical CPS applications include a variety of hardware types with different capabilities in power, computation and storage, the trust and reputation model should take into account these conditions of heterogeneity~\cite{Sharma2020}. For instance, when the trust score is derived from the computational power of a device, the model should apply an appropriate weighting for high-performance and constrained devices to overcome the heterogeneity in computational power.

\section{Conclusion}
\label{sec:conclusion}
The goal of a Trust and Reputation Management Systems (TRMS) is to assess and quantify the trustworthiness of each participant in the system for safeguarding users from the risk of interacting with untrustworthy entities. Adoption of blockchain to TRMS can benefit and enhance TRMS. In this article, we presented how the salient features of blockchain can enhance TRMS. We specifically focused on four features of blockchain, namely decentralisation, smart contracts, pseudonyms, immutable storage and transparent mechanism. We described several implementations of DTRMS across different fields of applications. We also discussed several open challenges and future directions for research in the area of DTRMS.

\section*{Acknowledgement}
This work was supported by the Cyber Security Research Centre Limited through the Australian Government’s Cooperative Research Centres Programme.

\printbibliography

@article{Bai2021,
author          = {Bai, Yuhan and Fan, Kai and Zhang, Kuan and Cheng, Xiaochun and Li, Hui and Yang, Yintang},
doi             = {10.1016/j.jclepro.2021.127407},
issn            = {09596526},
journal         = {Journal of Cleaner Production},
month           = {5},
pages           = {127407},
publisher       = {Elsevier},
title           = {{Blockchain-based trust management for agricultural green supply: A game theoretic approach}},
year            = {2021}
}

@Inbook{Dedeoglu2020,
author          = "Dedeoglu, V. and Jurdak, R. and Dorri, A. and Lunardi, R. C. and Michelin, R. A. and Zorzo, A. F. and Kanhere, S. S.",
editor          = "Kim, Shiho
and Deka, Ganesh Chandra",
title           = "Blockchain Technologies for IoT",
bookTitle       = "Advanced Applications of Blockchain Technology",
year="2020",
publisher       = "Springer Singapore",
address         = "Singapore",
pages           = "55--89",
doi             = "10.1007/978-981-13-8775-3_3",
url             = "https://doi.org/10.1007/978-981-13-8775-3_3"
}

@article{Feng2019,
author          = {Feng, Wei and Yan, Zheng},
doi             = {10.1016/j.future.2019.01.036},
issn            = {0167739X},
journal         = {Future Generation Computer Systems},
pages           = {649--666},
publisher       = {Elsevier B.V.},
title           = {{MCS-Chain: Decentralized and trustworthy mobile crowdsourcing based on blockchain}},
url             = {https://doi.org/10.1016/j.future.2019.01.036},
volume          = {95},
year            = {2019}
}

@article{Yan2021,
author          = {Yan, Zheng and Peng, Li and Feng, Wei and Yang, Laurence T.},
doi             = {10.1145/3419102},
issn            = {15576051},
journal         = {ACM Transactions on Internet Technology},
number          = {1},
title           = {{Social-Chain: Decentralized Trust Evaluation Based on Blockchain in Pervasive Social Networking}},
volume          = {21},
year            = {2021}
}

@INPROCEEDINGS{sidra-trustchain,
author          = {Malik, Sidra and Dedeoglu, Volkan and Kanhere, Salil S. and Jurdak, Raja},
booktitle       = {2019 IEEE International Conference on Blockchain (Blockchain)}, 
title           = {TrustChain: Trust Management in Blockchain and IoT Supported Supply Chains}, 
year            = {2019},
volume          = {},
number          = {},
pages           = {184-193},
doi             = {10.1109/Blockchain.2019.00032}
}

@article{Batwa2021,
author          = {Batwa, Abbas and Norrman, Andreas},
title           = {Blockchain Technology and Trust in Supply Chain Management: A Literature Review and Research Agenda},
journal         = {Operations and Supply Chain Management: An International Journal},
volume          = {14},
number          = {2},
pages           = {203--220},
year            = {2021},
publisher       = {OSCM Forum},
doi             = {http://doi.org/10.31387/oscm0450297}
}

@article{Zhao2021,
author          = {Zhao, Yanan and Wang, Yunpeng and Wang, Pengcheng and Yu, Haiyang},
doi             = {10.1109/JSYST.2021.3078797},
issn            = {1932-8184},
journal         = {IEEE Systems Journal},
pages           = {1--10},
title           = {{PBTM: A Privacy-Preserving Announcement Protocol With Blockchain-Based Trust Management for IoV}},
url             = {https://ieeexplore.ieee.org/document/9442949/},
year            = {2021}
}

@ARTICLE{putra-tnsm,
author          = {Putra, Guntur Dharma and Dedeoglu, Volkan and Kanhere, Salil S and Jurdak, Raja and Ignjatovic, Aleksandar},
journal         = {IEEE Transactions on Network and Service Management}, 
title           = {Trust-based Blockchain Authorization for IoT}, 
year            = {2021},
volume          = {},
number          = {},
pages           = {1-1},
doi             = {10.1109/TNSM.2021.3077276}
}

@INPROCEEDINGS{putra-icbc2020,
author          = {Putra, Guntur Dharma and Dedeoglu, Volkan and Kanhere, Salil S. and Jurdak, Raja},
booktitle       = {2020 IEEE International Conference on Blockchain and Cryptocurrency (ICBC)}, 
title           = {Trust Management in Decentralized IoT Access Control System}, 
year            = {2020},
volume          = {},
number          = {},
pages           = {1-9},
doi             = {10.1109/ICBC48266.2020.9169481}
}

@ARTICLE{kouicem,
author          = {Kouicem, Djamel Eddine and Imine, Youcef and Bouabdallah, Abdelmadjid and Lakhlef, Hicham},
journal         = {IEEE Transactions on Dependable and Secure Computing}, 
title           = {A Decentralized Blockchain-Based Trust Management Protocol for the Internet of Things}, 
year            = {2020},
volume          = {},
number          = {},
pages           = {1-1},
doi             = {10.1109/TDSC.2020.3003232}
}

@book{Li2020,
author          = {Li, Haochen and Gai, Keke and Zhu, Liehuang and Jiang, Peng and Qiu, Meikang},
booktitle       = {Lecture Notes in Computer Science (including subseries Lecture Notes in Artificial Intelligence and Lecture Notes in Bioinformatics)},
doi             = {10.1007/978-3-030-60248-2_3},
isbn            = {9783030602475},
issn            = {16113349},
pages           = {35--49},
publisher       = {Springer International Publishing},
title           = {{Reputation-Based Trustworthy Supply Chain Management Using Smart Contract}},
url             = {http://dx.doi.org/10.1007/978-3-030-60248-2_3},
volume          = {12454 LNCS},
year            = {2020}
}

@article{Wang2019,
author          = {Wang, Yuntao and Su, Zhou and Zhang, Ning},
doi             = {10.1109/TII.2019.2908497},
issn            = {15513203},
journal         = {IEEE Transactions on Industrial Informatics},
number          = {6},
pages           = {3620--3631},
publisher       = {IEEE},
title           = {{BSIS: Blockchain-based secure incentive scheme for energy delivery in vehicular energy network}},
volume          = {15},
year            = {2019}
}

@article{Yang2019,
author          = {Yang, Zhe and Yang, Kan and Lei, Lei and Zheng, Kan and Leung, Victor C.M.},
doi             = {10.1109/JIOT.2018.2836144},
issn            = {23274662},
journal         = {IEEE Internet of Things Journal},
number          = {2},
pages           = {1495--1505},
publisher       = {IEEE},
title           = {{Blockchain-based decentralized trust management in vehicular networks}},
volume          = {6},
year            = {2019}
}

@INPROCEEDINGS{camilo2020,
author          = {Camilo, Gustavo F. and Rebello, Gabriel Antonio F. and de Souza, Lucas Airam C. and Duarte, Otto Carlos M. B.},
booktitle       = {2020 IEEE International Conference on Blockchain (Blockchain)}, 
title           = {A Secure Personal-Data Trading System Based on Blockchain, Trust, and Reputation}, 
year            = {2020},
volume          = {},
number          = {},
pages           = {379-384},
doi             = {10.1109/Blockchain50366.2020.00055}
}

@InProceedings{javaid2020,
author          = "Javaid, Atia and Zahid, Maheen and Ali, Ishtiaq and Khan, Raja Jalees Ul Hussen and Noshad, Zainib and Javaid, Nadeem",
editor          = "Barolli, Leonard and Hellinckx, Peter and Enokido, Tomoya",
title           = "Reputation System for IoT Data Monetization Using Blockchain",
booktitle       = "Advances on Broad-Band Wireless Computing, Communication and Applications",
year            = "2020",
publisher       = "Springer International Publishing",
address         = "Cham",
pages           = "173--184",
isbn            = "978-3-030-33506-9"
}

@INPROCEEDINGS{sonbol-icbc2021,
author          = {Sonbol Rahimpour and Majid Khabbazian},
booktitle       = {2021 IEEE International Conference on Blockchain and Cryptocurrency (ICBC)}, 
title           = {Hashcashed Reputation with Application in Designing Watchtowers}, 
year            = {2021},
volume          = {},
number          = {},
pages           = {1-9}
}

@article{Zhao2018,
author          = {Zhao, Yanqi and Li, Yannan and Mu, Qilin and Yang, Bo and Yu, Yong},
doi             = {10.1109/ACCESS.2018.2799205},
issn            = {21693536},
journal         = {IEEE Access},
pages           = {12295--12303},
publisher       = {IEEE},
title           = {{Secure Pub-Sub: Blockchain-Based Fair Payment with Reputation for Reliable Cyber Physical Systems}},
volume          = {6},
year            = {2018}
}

@inproceedings{alowayed2018,
author          = {Alowayed, Yousef and Canini, Marco and Marcos, Pedro and Chiesa, Marco and Barcellos, Marinho},
title           = {Picking a Partner: A Fair Blockchain Based Scoring Protocol for Autonomous Systems},
year            = {2018},
isbn            = {9781450355858},
publisher       = {Association for Computing Machinery},
address         = {New York, NY, USA},
url             = {https://doi.org/10.1145/3232755.3232785},
doi             = {10.1145/3232755.3232785},
booktitle       = {Proceedings of the Applied Networking Research Workshop},
pages           = {33–39},
numpages        = {7},
location        = {Montreal, QC, Canada},
series          = {ANRW '18}
}

@inproceedings{dorigo2018blockchain,
title           = {Blockchain Technology for Robot Swarms: A Shared Knowledge and Reputation Management System for Collective Estimation},
author          = {Strobel, Volker and Dorigo, Marco},
booktitle       = {Swarm Intelligence--Proceedings of ANTS 2018--Eleventh International Conference},
pages           = {425--426},
year            = {2018},
organization    = {Springer}
}

@article{Lu2018,
archivePrefix   = {arXiv},
arxivId         = {1807.06159},
author          = {Lu, Zhaojun and Wang, Qian and Qu, Gang and Liu, Zhenglin},
doi             = {10.1109/TrustCom/BigDataSE.2018.00025},
eprint          = {1807.06159},
isbn            = {9781538643877},
issn            = {2324-9013},
journal         = {Proceedings - 17th IEEE International Conference on Trust, Security and Privacy in Computing and Communications and 12th IEEE International Conference on Big Data Science and Engineering, Trustcom/BigDataSE 2018},
pages           = {98--103},
publisher       = {IEEE},
title           = {{BARS: A Blockchain-Based Anonymous Reputation System for Trust Management in VANETs}},
year            = {2018}
}

@article{Yahaya2020,
author          = {Yahaya, Adamu Sani and Javaid, Nadeem and Javed, Muhammad Umar and Shafiq, Muhammad and Khan, Wazir Zada and Aalsalem, Mohammed Y.},
doi             = {10.1109/ACCESS.2020.3041931},
issn            = {21693536},
journal         = {IEEE Access},
pages           = {222168--222186},
title           = {{Blockchain-Based Energy Trading and Load Balancing Using Contract Theory and Reputation in a Smart Community}},
volume          = {8},
year            = {2020}
}

@article{Khorasany2021,
archivePrefix   = {arXiv},
arxivId         = {2005.14520},
author          = {Khorasany, Mohsen and Dorri, Ali and Razzaghi, Reza and Jurdak, Raja},
doi             = {10.1016/j.ijepes.2020.106610},
eprint          = {2005.14520},
issn            = {01420615},
journal         = {International Journal of Electrical Power and Energy Systems},
number          = {October 2020},
pages           = {106610},
publisher       = {Elsevier Ltd},
title           = {{Lightweight blockchain framework for location-aware peer-to-peer energy trading}},
url             = {https://doi.org/10.1016/j.ijepes.2020.106610},
volume          = {127},
year            = {2021}
}

@article{Wang2021,
author          = {Wang, Tonghe and Guo, Jian and Ai, Songpu and Cao, Junwei},
doi             = {10.1016/j.apenergy.2021.117056},
issn            = {03062619},
journal         = {Applied Energy},
number          = {May},
publisher       = {Elsevier Ltd},
title           = {{RBT: A distributed reputation system for blockchain-based peer-to-peer energy trading with fairness consideration}},
volume          = {295},
year            = {2021}
}

@ARTICLE{repucoin,
author          = {Yu, Jiangshan and Kozhaya, David and Decouchant, Jeremie and Esteves-Verissimo, Paulo},
journal         = {IEEE Transactions on Computers}, 
title           = {RepuCoin: Your Reputation Is Your Power}, 
year            = {2019},
volume          = {68},
number          = {8},
pages           = {1225-1237},
doi             = {10.1109/TC.2019.2900648}
}

@article{gambetta2000can,
title           = {Can we trust trust},
author          = {Gambetta, Diego and others},
journal         = {Trust: Making and breaking cooperative relations},
volume          = {13},
pages           = {213--237},
year            = {2000}
}

@ARTICLE{Zou2019,
author          = {Zou, Jun and Ye, Bin and Qu, Lie and Wang, Yan and Orgun, Mehmet A. and Li, Lei},
journal         = {IEEE Transactions on Services Computing}, 
title           = {A Proof-of-Trust Consensus Protocol for Enhancing Accountability in Crowdsourcing Services}, 
year            = {2019},
volume          = {12},
number          = {3},
pages           = {429-445},
doi             = {10.1109/TSC.2018.2823705}
}

@inproceedings{Dedeoglu2019,
author          = {Dedeoglu, Volkan and Jurdak, Raja and Putra, Guntur D. and Dorri, Ali and Kanhere, Salil S.},
title           = {A Trust Architecture for Blockchain in IoT},
year            = {2019},
isbn            = {9781450372831},
publisher       = {Association for Computing Machinery},
address         = {New York, NY, USA},
url             = {https://doi.org/10.1145/3360774.3360822},
doi             = {10.1145/3360774.3360822},
booktitle       = {Proceedings of the 16th EAI International Conference on Mobile and Ubiquitous Systems: Computing, Networking and Services},
pages           = {190–199},
numpages        = {10},
location        = {Houston, Texas, USA},
series          = {MobiQuitous '19}
}

@article{Bellini2020,
author          = {Bellini, Emanuele and Iraqi, Youssef and Damiani, Ernesto},
doi             = {10.1109/ACCESS.2020.2969820},
issn            = {21693536},
journal         = {IEEE Access},
pages           = {21127--21151},
title           = {{Blockchain-Based Distributed Trust and Reputation Management Systems: A Survey}},
volume          = {8},
year            = {2020}
}

@article{Hasan2020,
author          = {Hasan, Omar and Brunie, Lionel and Bertino, Elisa},
journal         = {University of Lyon Research Report},
pages           = {1--65},
title           = {{Privacy Preserving Reputation Systems based on Blockchain and other Cryptographic Building Blocks: A Survey}},
year            = {2020}
}

@article{Sharma2020,
author          = {Sharma, Avani and Pilli, Emmanuel S. and Mazumdar, Arka P. and Gera, Poonam},
doi             = {10.1016/j.comcom.2020.06.030},
issn            = {1873703X},
journal         = {Computer Communications},
number          = {June},
pages           = {475--493},
publisher       = {Elsevier B.V.},
title           = {{Towards trustworthy Internet of Things: A survey on Trust Management applications and schemes}},
url             = {https://doi.org/10.1016/j.comcom.2020.06.030},
volume          = {160},
year            = {2020}
}

@INPROCEEDINGS{Clemence2019,
author          = {Dorri, Ali and Roulin, Clemence and Jurdak, Raja and Kanhere, Salil S.},
booktitle       = {2019 IEEE 44th Conference on Local Computer Networks (LCN)}, 
title={On the Activity Privacy of Blockchain for IoT}, 
year            = {2019},
volume          = {},
number          = {},
pages           = {258-261},
doi             = {10.1109/LCN44214.2019.8990819}
}

@article{Chahal2020,
author          = {Chahal, Rajanpreet Kaur and Kumar, Neeraj and Batra, Shalini},
doi             = {10.1016/j.comcom.2019.10.034},
issn            = {1873703X},
journal         = {Computer Communications},
number          = {November 2019},
pages           = {13--46},
publisher       = {Elsevier B.V.},
title           = {{Trust management in social Internet of Things: A taxonomy, open issues, and challenges}},
url             = {https://doi.org/10.1016/j.comcom.2019.10.034},
volume          = {150},
year            = {2020}
}

@inproceedings{boldyreva2009order,
title           = {Order-preserving symmetric encryption},
author          = {Boldyreva, Alexandra and Chenette, Nathan and Lee, Younho and O’neill, Adam},
booktitle       = {Annual International Conference on the Theory and Applications of Cryptographic Techniques},
pages           = {224--241},
year            = {2009},
organization    = {Springer}
}

@article{back2002hashcash,
title           = {Hashcash-A Denial of Service Counter-Measure},
author          = {Back, Adam},
year            = {2002}
}

@misc{poon2016bitcoin,
title           = {The bitcoin lightning network: Scalable off-chain instant payments},
author          = {Poon, Joseph and Dryja, Thaddeus},
year            = {2016}
}

@article{goldwasser1989knowledge,
title           = {The knowledge complexity of interactive proof systems},
author          = {Goldwasser, Shafi and Micali, Silvio and Rackoff, Charles},
journal         = {SIAM Journal on computing},
volume          = {18},
number          = {1},
pages           = {186--208},
year            = {1989},
publisher       = {SIAM}
}

@inproceedings{douceur2002sybil,
title           = {The sybil attack},
author          = {Douceur, John R},
booktitle       = {International workshop on peer-to-peer systems},
pages           = {251--260},
year            = {2002},
organization    = {Springer}
}

@inproceedings{yu2008sybillimit,
title           = {Sybillimit: A near-optimal social network defense against sybil attacks},
author          = {Yu, Haifeng and Gibbons, Phillip B and Kaminsky, Michael and Xiao, Feng},
booktitle       = {2008 IEEE Symposium on Security and Privacy (sp 2008)},
pages           = {3--17},
year            = {2008},
organization    = {IEEE}
}

@inproceedings{josang2009challenges,
title           = {Challenges for robust trust and reputation systems},
author          = {J{\o}sang, Audun and Golbeck, Jennifer},
booktitle       = {Proceedings of the 5th International Workshop on Security and Trust Management (SMT 2009), Saint Malo, France},
volume          = {5},
number          = {9},
year            = {2009},
organization    = {Citeseer}
}

@article{hoffman2009survey,
title           = {A survey of attack and defense techniques for reputation systems},
author          = {Hoffman, Kevin and Zage, David and Nita-Rotaru, Cristina},
journal         = {ACM Computing Surveys (CSUR)},
volume          = {42},
number          = {1},
pages           = {1--31},
year            = {2009},
publisher       = {ACM New York, NY, USA}
}

@article{stoll2019carbon,
title           = {The carbon footprint of bitcoin},
author          = {Stoll, Christian and Klaa{\ss}en, Lena and Gallersd{\"o}rfer, Ulrich},
journal         = {Joule},
volume          = {3},
number          = {7},
pages           = {1647--1661},
year            = {2019},
publisher       = {Elsevier}
}

@misc{dorri2016blockchain,
title           = {Blockchain in internet of things: Challenges and Solutions}, 
author          = {Ali Dorri and Salil S. Kanhere and Raja Jurdak},
year            = {2016},
eprint          = {1608.05187},
archivePrefix   = {arXiv},
primaryClass    = {cs.CR}
}

@article{dorri2019lsb,
title           = {LSB: A Lightweight Scalable Blockchain for IoT security and anonymity},
author          = {Dorri, Ali and Kanhere, Salil S and Jurdak, Raja and Gauravaram, Praveen},
journal         = {Journal of Parallel and Distributed Computing},
volume          = {134},
pages           = {180--197},
year            = {2019},
publisher       = {Elsevier}
}

@article{josang2007survey,
title           = {A survey of trust and reputation systems for online service provision},
author          = {J{\o}sang, Audun and Ismail, Roslan and Boyd, Colin},
journal         = {Decision support systems},
volume          = {43},
number          = {2},
pages           = {618--644},
year            = {2007},
publisher       = {Elsevier}
}

@article{Chen2016,
author          = {Chen, Ing Ray and Bao, Fenye and Guo, Jia},
doi             = {10.1109/TDSC.2015.2420552},
issn            = {19410018},
journal         = {IEEE Transactions on Dependable and Secure Computing},
number          = {6},
pages           = {684--696},
publisher       = {IEEE},
title           = {{Trust-Based Service Management for Social Internet of Things Systems}},
volume          = {13},
year            = {2016}
}

@inproceedings{levien1998attack,
title           = {Attack-Resistant Trust Metrics for Public Key Certification.},
author          = {Levien, Raph and Aiken, Alex},
booktitle       = {Usenix security symposium},
pages           = {229--242},
year            = {1998}
}

@inproceedings{kamvar2003eigentrust,
title           = {The eigentrust algorithm for reputation management in p2p networks},
author          = {Kamvar, Sepandar D and Schlosser, Mario T and Garcia-Molina, Hector},
booktitle       = {Proceedings of the 12th international conference on World Wide Web},
pages           = {640--651},
year            = {2003}
}

@inproceedings{whitby2004filtering,
title           = {Filtering out unfair ratings in bayesian reputation systems},
author          = {Whitby, Andrew and J{\o}sang, Audun and Indulska, Jadwiga},
booktitle       = {Proc. 7th Int. Workshop on Trust in Agent Societies},
volume          = {6},
pages           = {106--117},
year            = {2004},
organization    = {Citeseer}
}

\end{document}